\documentclass[12pt]{article}
\usepackage{amsfonts}
\usepackage{amsmath}
\usepackage{amssymb}
\usepackage{graphicx}
\usepackage{subfigure}
\usepackage{ccaption}

\captiondelim{. }

\def \be {\begin{equation}}
\def \ee {\end{equation}}
\def \bea {\begin{eqnarray}}
\def \eea {\end{eqnarray}}
\def \nn {\nonumber}

\def \rr {\raise.35ex\hbox{\small $\prime$}\kern-.17em{\mbox{\large $\imath$}}}
\def \del {\partial}
\def \dels {\partial\kern-.5em / \kern.5em}
\def \As {{A\kern-.5em / \kern.5em}}
\def \Ds {D\kern-.7em / \kern.5em}
\def \ks {k\kern-.6em / \kern.5em}
\def \slash {\kern-.5em / \kern .5em}

\def \a {\alpha}

\def \Lam {\Lambda}

\def \Tr {\mathrm{Tr}}

\begin{document}
\begin{titlepage}
%\catcode`\@=11
%\catcode`\@=12
%\twocolumn[\hsize\textwidth\columnwidth\hsize\csname%
%@twocolumnfalse\endcsname

%\draft
\begin{center}
\hfill hep-th/yymmnnn\\
\vskip .5in

\textbf{\Large
A UV completion of scalar field theory \\
in arbitrary even dimensions}
\vskip .5in
{\large Pei-Ming Ho, Xue-Yan Lin}

\vskip 15pt

{\small Department of Physics, 
Center for Theoretical Sciences \\
National Taiwan University, Taipei 10617, Taiwan,
R.O.C.}

\vskip .2in
\sffamily{
pmho@phys.ntu.edu.tw \\
xueyan.lin@msa.hinet.net}

\vspace{60pt}
%\maketitle
\end{center}

\begin{abstract}

Following a previous work (hep-th/0410248),
where a scalar field theory with a modified propagator 
and $\phi^4$ interaction in 4 dimensions
is constructed to be UV-finite, unitary and Lorentz invariant, 
we discuss in this paper general $\phi^n$ theory 
in arbitrary even space-time dimensions. 
We show that the theory is still 
UV-finite, unitary and Lorentz invariant 
if the propagators are chosen to meet 
certain simple conditions depending on the space-time dimension 
but independent of $n$.
We also comment that our model is reminiscent of string theory 
in the way UV divergence is avoided.

\end{abstract}

{\center

{\em This paper is dedicated to the memory of Yi-Ya Tian.}

}

%\pacs{PACS numbers: 11.25.-w, 11.25.Mj, 11.25.Sq}%]
\end{titlepage}
%\begin{narrowtext}
\setcounter{footnote}{0}

\section{Introduction}

UV divergences in quantum field theories can be 
regularized by introducing higher derivatives 
in the kinetic term so that the propagator 
approaches faster to $0$ than $1/k^2$ at large momenta $k$. 
But this is usually done at the cost of unitarity. 
For example, by the Pauli-Villars regularization \cite{2,3} 
the propagator is modified as
\begin{equation}
\frac{1}{k^2+m^2}\longrightarrow\frac{1}{(k^2+m^2)(k^2+M^2)}
=\frac{1}{M^2-m^2}\left(
\frac{1}{k^2+m^2}-\frac{1}{k^2+M^2}
                  \right).
\end{equation} 
This propagator $\sim 1/k^4$ at large $k$, 
hence alleviates the UV divergence of a Feynman diagram. 
However, the norm of the propagating mode 
at $k^2 = -M^2$ is negative 
due to the minus sign of the pole at $M^2$. 
Unitarity is violated for energies beyond the ghost mass $M$.

In \cite{1}, a higher derivative correction to the propagator
of the form 
\begin{equation}
f(k^2)=\sum_{n=0}^{\infty} 
\frac{c_n}{k^2+m_n^2},\quad\quad c_n>0\quad \forall n 
\label{propagator}
\end{equation}
is considered.
Due to the condition $c_n > 0$, 
Cutkosky's rules \cite{4} ensure  
purturbative unitarity for generic Feynman diagrams. 
\footnote{
In the check of unitarity, only poles with masses 
lower than the center of mass energy $E$ need to be considered, 
and thus the fact that there are infinitely many poles 
in the propagator is irrelevant. 
}
In this paper we adopt the same type of propagators, 
thus our theories are automatically unitary. 
The models we consider are also manifestly Lorentz invariant. 
Hence, in the following
we will focus our attention on the removal of UV divergence.

In \cite{1} it was shown that 
to avoid UV divergence in the four dimensional $\phi^4$ theory, 
the following conditions are sufficient:
\begin{subequations}
\label{condition}
\begin{align}
&\sum_n c_n m_n^2=0, \\
&\sum_n c_n=0.
\end{align}
\end{subequations}
Since all the parameters $c_n$'must be greater than zero 
to ensure unitarity, 
thes conditions look impossible. 
The trick is that, since there is an infinite number of $c_n$'s, 
analytic continuation can be used \cite{1} to 
satisfy both conditions in eq. (\ref{condition}).

For example, with two constant parameters $z > 0$ and $a > 0$, 
let 
\begin{subequations}
\begin{align}
c_0&=\frac{1}{1-e^{-z}},\\
c_n&=e^{zn}\qquad n\ge 1,\\
m^2_0&=\frac{1-e^{-z}}{1-e^{-(z+a)}},\\
m^2_n&=e^{an},\qquad n\ge 1, 
\end{align}
\end{subequations}
then one can analytically continue the infinite sum 
to a simple form
\begin{subequations}
\label{zeta}
\begin{align}
\sum_{n=0}^{\infty} c_n &= 
\frac{1}{1-e^{-z}} + \sum_{n=1}^\infty e^{zn}
=\frac{1}{1-e^{-z}} + \frac{e^z}{1-e^z} = 0, \\
\sum_{n=0}^{\infty} c_n m_n^2 
&= \frac{1}{1-e^{-(z+a)}} + \sum_{n=1}^{\infty} e^{n(z+a)} 
= \frac{1}{1-e^{-(z+a)}} + \frac{e^{z+a}}{1-e^{z+a}} = 0.
\end{align}
\label{examplealc}
\end{subequations}
Note that the two infinite series above diverge 
if $z > 0$ or $(z+a) > 0$, respectively,  
but we define them by analytic continuation. 

Three important issues must be addressed immediately. 
First, it is important that the analytic continuation can 
be carried out consistently throughout all calculations. 
This will be the main concern when we give a prescription 
for the computation of Feynman diagrams. 

Secondly, some of the readers may be uncomfortable with 
this analytic continuation, 
in the absence of an intuitive physical interpretation.
However, we will point out that 
a similar analytic continuation is naturally incorporated in string theory 
from the viewpoint of the worldsheet theory.
It will be very interesting to construct 
an analogous worldsheet theory that will 
directly justify the analytic continuation used in our models.
But we shall leave this problem for the future.

Finally, while there is an infinite number of poles 
in the propagator,
this theory is also equivalent to a theory with 
an infinite number of scalar fields with masses $m_n$. 
If $m^2_n \gg m^2_0$ for all $n > 0$, 
the low energy behavior of this theory 
is approximated by an ordinary scalar field theory 
with a single scalar field with mass $m_0$. 

In this paper we will discuss a generic scalar field theory 
in general even dimensions. 
After studying the relations among interaction vertices, internal lines, 
external lines and loops in Feynman diagrams, 
we enumerate the conditions sufficient to eliminate 
all superficial divergences to ensure UV-finiteness. 
In the last section, we will discuss the physical meaning of 
analytic continuation, making an analogy with string theory.

\section{$\phi^n$ theory in 4 dimensions}

In this section we study the patterns of UV-divergence 
in a $\phi^n$ theory in 4 dimensional space-time, 
and list all the conditions needed to eliminate all the UV-divergences.
Roughly speaking, the more divergent a Feynman diagram is, 
the more conditions we need to make it finite. 
Thus we are particularly interested in the most divergent diagrams 
in order to find all the conditions needed to guarantee UV finiteness.
For the sake of simplicity, 
we assume that there is a unique $\phi^n$ interaction in the theory. 
Nevertheless, our conclusion will also apply to more general theories
including $\phi^{n-2}, \cdots, \phi^4$ interactions, 
since one can always construct the most divergent diagrams with 
$\phi^n$ interactions alone. 

In a diagram with superficial divergence of dimension $D$, 
in general there are divergent terms proportional to \cite{1}
\begin{equation}
\sum_n c_n \Lam^D,\quad\sum_nc_nm_n^2\Lam^{D-2},
\quad \cdots, \quad
\sum_nc_nm_n^{D-2}\Lam^{2},
\quad\sum_nc_nm_n^{D}\log(\Lam^2),
\end{equation} 
In 4 dimensions, 
the superficial divergence $D$ is determined by 
the number of loops $L$ and 
the number of internal lines (propagators) $I$ as
\begin{equation}
D=4L-2I.
\label{DLI}
\end{equation}
On the other hand, the number of loops $L$
is related to the number of vertices $V$ and internal lines $I$ via
\footnote{
This equality does not apply to the one-loop diagram 
without vertices ($I = 1$, $V = 0$ and $L = 1$).
This is because the propagator in the loop does not 
have its endpoints ending on vertices.
}
\begin{equation}
L=I-V+1.
\label{loopandvertex}
\end{equation}
This equation can be understood as follows.
The calculation of a Feynman diagram with $L$ loops 
always turns out to be an integration over 
$L$ free momentum parameters ($p_1, \cdots, p_L$).
On the other hand, 
the number of free momentum parameters 
should also equal the total number of momenta $I$
assigned to each propagator ($q_1, \cdots, q_I$) 
minus the number of constraints $V$ 
for the momentum conservation at each vertex. 
However, the constraints of momentum conservation 
at all vertices are not linearly independent. 
The number $1$ on the right hand side of (\ref{loopandvertex})
corresponds to the momentum conservation 
of the whole diagram, 
which is automatically satisfied by the assignment of external momenta. 

Another equality that will be used later is
\begin{equation}
E = nV - 2I,
\label{EVI}
\end{equation}
where  $E$ is the number of external lines
and $n$ the number of legs of each interaction vertex.
Using the relations above, we can express $D$ as
\begin{equation}
D=(n-4)V-E+4.
\end{equation}

In the 4 dimensional $\phi^4$ theory, 
$D$ only depends on the number of external lines $E$ as $D = 4-E$, 
and is thus bounded from above by $D\leq 4$.
This is why we only need two conditions (\ref{condition})
to eliminate the divergences of $D = 4$ and $D = 2$. 

\begin{figure}
\begin{center}
\includegraphics[scale=0.8]{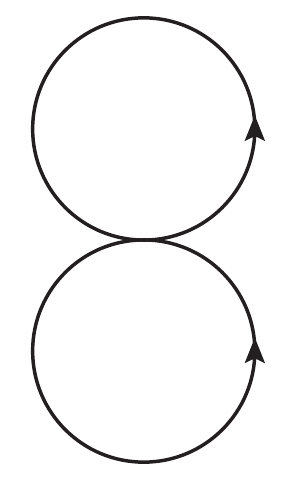}
\end{center}
\caption{}
\label{scalartwoloops}
\end{figure}

For $\phi^n$ theories with $n>4$, 
the large $V$ is, the higher superficial divergence $D$ can be. 
{\em A priori} this may enforce us to impose infinitely many conditions 
of the form 
\be
\sum_n c_n m_n^{2r}=0
\label{typicalcond}
\ee
with $r = 0, 1, 2, \cdots, \infty$.
However the divergence of a diagram can sometimes be decomposed 
into lower dimensional divergences.
For example, in the $\phi^4$ theory, 
there is a diagram with superficial divergence $D=4$
(see Fig.\ref{scalartwoloops}),
but since the two loops are separable, 
this diagram only needs a single condition of dimension $2$
(\ref{condition}a) to avoid the divergence. 

From section 3.4 of \cite{1}, 
a generic Feynman diagram with $L$ loops is of the form
\begin{equation}
\mathcal{M}=\sum_{n_1\cdots n_I}c_{n_1}\cdots c_{n_I}
\int d^4p_1\cdots \int d^4 p_L \prod_{i=1}^I\frac{1}{q_i^2+m_{n_i}^2},
\end{equation}
where $q_i$ is the momentum of the $i$-th internal line, 
which is a linear combination of the loop mementa $p_j$ 
and the momenta of external lines $k_i$. 
Using Feynman's parameters, this quantity can be rewritten as
\begin{equation}
\mathcal{M}\propto\sum_{n_1\cdots n_I}c_{n_1}\cdots c_{n_I}
\int_0^1 d\alpha_1 \cdots \int_0^1 d\alpha_I\delta(\a_1+\cdots+\a_I)
\int d^4p_1\cdots \int d^4 p_L 
\frac{1}{\left(\sum_{i=1}^I\a_i(q_i^2+m_{n_i}^2)\right)^I}.
\label{Mgeneral}
\end{equation}
By shifting the loop momenta $p_j\rightarrow p'_j$, the integrand can be simplified as
\begin{equation}
\frac{1}{\left(\sum_{j=1}^L\beta_jp_j'^2+\Delta\right)^I},
\end{equation}
where $\beta_j$'s are functions of the parameters $\a_i$, and 
\begin{equation}
\Delta=\Delta_0+\sum_{i,j=1}^E A_{ij}k_ik_j, 
\quad \Delta_0=\sum_{i=1}^I\a_im_{n_i}^2,
\label{DD}
\end{equation}
where $A_{ij}$'s are function of the Feynman parameters $\a_i$. 

Before summing over $n_1, \cdots, n_I$, 
each integral in (\ref{Mgeneral}) is potentially divergent. 
Our prescription of calculation is to first regularize all divergent integrals 
by dimensional regularization $d= 4-\epsilon$, 
and after imposing the conditions (\ref{condition}), 
we take the limit $\epsilon\rightarrow 0$ to obtain the final result. 
One could also apply other regularization schemes instead of 
dimensional regularization. 
It was shown in \cite{1} that, for the diagrams we computed explicitly,
various different regularization methods give exactly the same result. 
This may be a general feature of our models, 
although a rigorous proof is yet to be given.

By the general formula of dimensional regularization\cite{7} 
\begin{equation}
\int \frac{d^dl}{(2\pi)^d}\frac{1}{(l^2+\nu^2)^n}=\frac{1}{(4\pi)^{d/2}}\frac{\Gamma(n-\frac{d}{2})}{\Gamma(n)}\left(\frac{1}{\nu^2}\right)^{n-\frac{d}{2}},
\end{equation}
apart from the integration over $\a$'s, 
equation(\ref{Mgeneral}) can be integrated over loop momenta $p'_i$'s one by one as
%\begin{subequations}
%\begin{align}
\begin{eqnarray}
&&\sum_{n_1\cdots n_I}c_{n_1}\cdots c_{n_I}
\int d^d p_1\cdots \int d^d p_L 
\frac{1}{\left(\sum_{j=1}^L\beta_jp_j'^2+\Delta\right)^I} \nn \\
&\propto&\sum_{n_1\cdots n_I}c_{n_1}\cdots c_{n_I}
\int d^d p_1\cdots \int d^d p_{L-1} 
\frac{1}{\left(\sum_{j=1}^{L-1}\beta_jp_j'^2+\Delta\right)^{I-d/2}}
\Gamma\left(I-\frac{d}{2}\right) \nn \\
&\propto&\sum_{n_1\cdots n_I}c_{n_1}\cdots c_{n_I}
\int d^d p_1\cdots \int d^d p_{L-2} 
\frac{1}{\left(\sum_{j=1}^{L-2}\beta_jp_j'^2+\Delta\right)^{I-2d/2}}
\Gamma\left(I-\frac{d}{2}\right)
\frac{\Gamma(I-\frac{2d}{2})}{\Gamma(I-\frac{d}{2})}. \nn \\
%\end{align}
\label{Mgeneral2}
\end{eqnarray}
%\end{subequations}
It is easy to see that 
the Gamma function appearing in the denominator 
after integrating over a loop momentum always cancels
the numerator due to the previous integral. 
After we integrate over all loop momenta, 
the final result of (\ref{Mgeneral2}) is proportional to
%\begin{subequations}
%\begin{align}
\begin{eqnarray}
&&
\sum_{n_1\cdots n_I}c_{n_1}\cdots c_{n_I}
\left(\frac{1}{\Delta}\right)^{I-Ld/2}
\Gamma\left(I-\frac{Ld}{2}\right) \nn \\
&=&
\sum_{n_1\cdots n_I}c_{n_1}\cdots c_{n_I}
\left(\frac{1}{\Delta}\right)^{\frac{\epsilon L}{2}-\frac{D}{2}}
\Gamma\left(\frac{\epsilon L}{2}-\frac{D}{2}\right) \nn \\
&\approx&
\sum_{n_1\cdots n_I}c_{n_1}\cdots c_{n_I} \Delta^{D/2}
\left[1+\frac{\epsilon L}{2}\log{\left(\frac{1}{\Delta}\right)}
+{\cal O}(\epsilon^2)\right]
\frac{(-1)^{D/2}}{\left(\frac{D}{2}\right)!}
\left[\frac{2}{L\epsilon}+\left(-\gamma+\sum_{k=1}^{D/2}\frac{1}{k}\right)
+{\cal O}(\epsilon)\right] \nn \\
&\propto&
\sum_{n_1\cdots n_I}c_{n_1}\cdots c_{n_I} \Delta^{D/2}
\left[\frac{2}{L\epsilon}+
\log{\left(\frac{1}{\Delta}\right)}+
\left(-\gamma+\sum_{k=1}^{D/2}\frac{1}{k}\right)+{\cal O}(\epsilon)\right],
\label{Mgeneral3}
\end{eqnarray}
%\end{align}
%\end{subequations}
where $D$ is the superficial degree of divergence and $\epsilon=4-d$. 
The UV divergence of the diagram is summarized in 
the first term in (\ref{Mgeneral3}),
which diverges in the limit $\epsilon\rightarrow 0$.
To eliminate this UV divergence, we need 
\begin{equation}
\sum_{n_1\cdots n_I}c_{n_1}\cdots c_{n_I}\Delta^{D/2}=0.
\label{20}
\end{equation}
If this condition is satisfied, 
the third term also vanishes and 
the second term in (\ref{Mgeneral3}) 
contributes to the finite part of the amplitude
\begin{equation}
\mathcal{M}\propto\sum_{n_1\cdots n_I}c_{n_1}\cdots c_{n_I}
\int_0^1 d\alpha_1 \cdots \int_0^1 d\alpha_I\delta(\a_1+\cdots+\a_I)
\Delta^{D/2}\log{\left(\frac{1}{\Delta}\right)}.
\end{equation}

As we look at diagrams with 
higher and higher superficial divergence $D$, 
there is a chance of finding new conditions 
of the form (\ref{typicalcond}) with larger and larger values of $r$ 
in order for (\ref{20}) to remain valid.
To understand the precise connection between $D$ and the values of $r$,  
we decompose (\ref{20}) intro equations of the form (\ref{typicalcond}) 
with different values of $r$. 
But we only care about the largest value of $r$, $r_{max}$,
(or the largest power on the masses $m_n$), 
since all conditions of the form (\ref{typicalcond}) with $r < r_{max}$ 
are needs for all diagrams to be UV finite.
According to (\ref{DD}), eq. (\ref{20}) can be expanded 
(note that $D$ is always even, see (\ref{DLI})) as 
\be
\sum_{n_1\cdots n_I}c_{n_1}\cdots c_{n_I}
\left(\Delta_0^{D/2} + 
\frac{D}{2} \Delta_0^{D/2-1}\left(\sum_{i, j=1}^E A_{ij}k_i k_j\right)
+ C^{D/2}_2 \Delta_0^{D/2-2}\left(\sum_{i, j=1}^E A_{ij}k_i k_j\right)^2 
+ \cdots
\right),
\ee
where $C^{D/2}_2 = \frac{(D/2)(D/2-1)}{2}$, 
and the largest power on $m_n^2$ in (\ref{20}) resides in the term
\begin{equation}
\sum_{n_1\cdots n_I}c_{n_1}\cdots c_{n_I}\Delta_0^{D/2}
=\sum_{n_1\cdots n_I}c_{n_1}\cdots c_{n_I}
\left(\sum_{i=1}^I\a_im_{n_i}^2\right)^{D/2}=0.
\label{n1n2n3am}
\end{equation}
Apparently, the conditions (\ref{condition})
($\sum c_n=0$ and $\sum c_n m_n^2=0$) 
needed for the $\phi^4$ theory 
must also be needed for $\phi^n$ theory with $n > 4$.
Thus we can first remove all the terms in (\ref{n1n2n3am}) 
that already vanish due to these conditions.
This means that in the expansion of $(\sum_{i=1}^I\a_im_{n_i}^2)^{D/2}$,
we must be able to associate at least two factors of $m_{n_i}^2$ 
to each $c_{n_i}$ in order for a particular term to survive. 
However, 
since each term in the expansion of $(\sum_{i=1}^I\a_im_{n_i}^2)^{D/2}$ 
is a product of $D/2$ powers of $\a_im_{n_i}^2$, 
and there are $I$ possible values of the index $i$ on $c_{n_i}$ to check, 
it will not be possible to associate two or more factors of $m_{n_i}^2$ 
for all values of $i$ if
\begin{equation}
%I \ge D/2=\frac{4L-2I}{2}=2L-I,
2I > D/2=\frac{4L-2I}{2}=2L-I.
\end{equation}
As a result there will be no condition other than 
$\sum c_n=0$ and $\sum c_n m_n^2$=0 if
\begin{equation}
%I \ge L.
3I > 2L.
\end{equation}
Combining this with eq. (\ref{loopandvertex}) 
leads to a trivial condition
\begin{equation}
%V \ge 1.
V+I/2 > 1.
\label{Vge1}
\end{equation}
This condition is violated only by the one-loop diagram 
without vertex ($V=0$), 
which is already considered in the $\phi^4$ theory 
and vanishes under the conditions (\ref{condition}).
Thus we have proven that in 4 dimensions 
all $\phi^n$ theories are UV finite
if the propagator (\ref{propagator}) 
satisfies the conditions (\ref{condition}).

\section{$\phi^n$ theory in arbitrary even dimensions}

In general, the relation between the superficial divergence $D$ 
and space-time dimension $d$ is
\begin{equation}
D=dL-2I.
\label{DdLI}
\end{equation}
In this paper we restrict our disscussion to 
the cases of even dimensional space-time. 
The reason is that odd dimensions may lead to 
odd values of superficial divergence $D$, 
and $\Delta^{D/2}$ is no longer a polynomial of $\Delta_0$. 

\begin{figure}
\begin{center}
\includegraphics[scale=0.7]{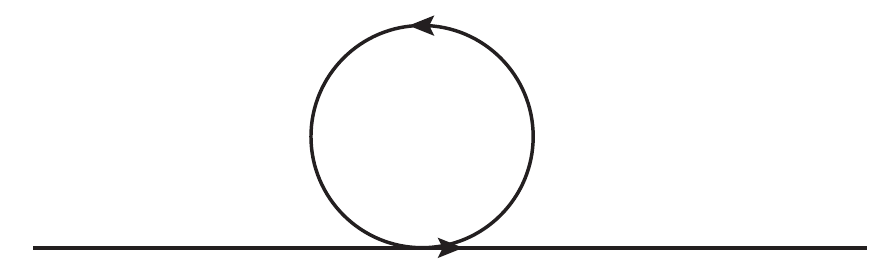}
\end{center}
\caption{}
\label{oneloop5D}
\end{figure}

Repeating the arguments in the previous section 
for a generic even dimension $d$, 
we find (\ref{Vge1}) replaced by 
\begin{equation}
%V \ge 1+\left(1-\frac4d\right)I.
V > 1+\left(1-\frac{6}{d}\right) I.
\end{equation}
This condition can be easily violated when $d > 4$. 
For example, the simple $\phi^4$ one-loop diagram
in Fig.\ref{oneloop5D} for 6 dimensional spacetime has 
a superficial divergence of $6\times 1-2\times 1=4$. 
Clearly we need one more condition $\sum c_n m_n^4=0$ 
in addition to (\ref{condition}).

The next question is: 
for given $n$ and $d>4$, 
do we need infinitely many conditions to ensure 
all diagrams to be finite, 
or only a finite number of conditions suffice to avoid all UV divergences?

To answer this question, 
we revisit eq. (\ref{n1n2n3am}) in more detail. 
If we impose sufficiently many conditions 
of the form (\ref{typicalcond}) to ensure 
that (\ref{n1n2n3am}) vanishes, 
there would be no UV divergences. 
If $2I\leq D/2$, 
there are terms in (\ref{n1n2n3am}) 
with a factor of $m_{n_i}$ to the 4th or higher powers 
associated with each factor of $c_{n_i}$'s,
and thus we need the condition $\sum c_n m_n^4=0$ 
in order to remove such terms. 
Similarly, if $3I\leq D/2$, 
we also need $\sum c_n m_n^6=0$, and so on. 
In general, for a Feynman diagram with superficial divergence $D$
and $I$ internal lines, 
we need conditions (\ref{typicalcond}) 
with $r = 1, 2, \cdots, [D/2I]$, 
where $[D/2I]$ denotes the integer part of $D/2I$.
Therefore, we are interested in the maximal value of 
$[D/2I]$ for a $\phi^n$ theory in $d$ dimensions with given $n$ and $d$.
%\begin{equation}
%rI\leq\frac D2, 
%\end{equation}
If the set of $[D/2I]$ for all Feynman diagrams 
is unbounded from above, 
we need an infinite number of conditions.

Using (\ref{DdLI}), and then (\ref{loopandvertex}), 
one can express $D$ in terms of $V$ and $I$ as 
\footnote{
Again we are excluding the diagram without vertices. 
It can be checked separately that 
the conditions we will impose later will also ensure 
that these diagrams are free of UV divergences.
}
\be
D = dL - 2I = d(I-V+1) - 2I = (d-2)I - d(V-1) \leq (d-2)I.
\ee
This implies that there is an upper bound to the number $[D/2I]$, i.e.
\be
D/2I \leq \frac{d-2}{2}.
\label{choiceofk}
\ee
This means that for any given $n$ and space-time dimension $d$, 
we only need the conditions 
\be
c_n m_n^{2r}=0 \qquad \mbox{for} \qquad r = 0, 1, \cdots, \frac{d-2}{2}.
\label{generalcond}
\ee
Remarkably, this condition is independent of $n$. 
It follows that, for given dimension $d$, 
the same propagator that satisfies (\ref{generalcond}) 
suits all polynomial interactions of $\phi$.
sion in our theory.

As it was commented in \cite{1}, 
the desired propagators satisfying all the conditions 
are easy to construct. 
Here we give a systematic way to construct 
propagators satisfying (\ref{generalcond}) for generic $d$.

With a set of $d/2$ positive parameters $x_i$, 
we define
\begin{subequations}
\label{example}
\begin{align}
c_n&=\left[
1+x_1(n+1)+x_2(n+2)(n+1)+\cdots 
+x_{d/2}\frac{(n+d/2)!}{n!}
\right]e^{zn}, \\
m^2_n&=e^{an}
\end{align}
for $n = 0, 1, 2, \cdots$.
\end{subequations}
Denoting $\rho\equiv e^{z+ar}$ for convenience, 
we carry out the infinite sum $\sum c_nm^{2r}_n$
first assuming that $\rho<1$, 
and then we analytically continue it back to $\rho>1$. 
The result of $\sum c_nm^{2r}_n$ is
\begin{subequations}
\begin{align}
\sum_{n=0}^\infty c_nm_n^{2r}&=\frac{1}{1-\rho}+
x_1\frac{d}{d\rho}\left(\frac{1}{1-\rho}\right)+
x_2\frac{d^2}{d\rho^2}\left(\frac{1}{1-\rho}\right)+
\cdots x_{d/2}\frac{d^{\frac{d}{2}}}{d\rho^\frac{d}{2}}\left(\frac{1}{1-\rho}\right)\\
&=\frac{1}{\xi}+\frac{x_1}{\xi^2}+\frac{x_2}{\xi^3}+
\cdots\frac{x_{d/2}}{\xi^{d/2+1}} \equiv h(\xi),
\end{align}
\end{subequations} 
where $\xi\equiv\frac{1}{1-\rho}$, which is negative definite when $\rho>1$.
We have sufficient parameters $\{x_1,x_2\cdots x_\frac{d}{2}\}$ 
to fix the $d/2$ roots of $\xi$ at desired positions 
$\{-\xi_1,-\xi_2,\cdots,-\xi_{d/2}\}$ ($\xi_i$'s are positive). 
We can find the correspondence between $x_i$'s and $\xi_i$'s from
\begin{equation}
\xi^{d/2+1}h(\xi) = 
c(\xi+\xi_1)(\xi+\xi_1)\cdots(\xi+\xi_{d/2}),
\label{findx}
\end{equation}
where $c$ is an arbitrary real parameter.
Apparently all $x_i$'s are positive because the polynomial (\ref{findx})
has no negative coefficients.
As a result, all $c_n$'s are positive and unitarity is preserved.

\section{Analytic continuation and string theory}
\label{AC}

\subsection{Analytic continuation}

It might appear strange to some readers that 
the analytic continuation of a parameter in the propagator is used 
to eliminate UV divergences. 
What is the physical meaning of this analytic continuation?
We will try to give some hint to answering this question.

First, analytic continuation means the extension of the domain 
of a function $f(x)$ under the requirement of analyticity. 
For example, if we define $f(x)$ by the series
\begin{equation}
f(x)=1+x+x^2+x^3+\cdots=\sum_{n=0}^\infty x^n,
\label{seriessol}
\end{equation}
the domain of $f(x)$ should be restricted to $(-1,1)$ 
because the radius of convergence is $1$. 
However, we can extend the definition of $f(x)$
by analytic continuation to the whole complex plane 
except the point at $x=1$, so that 
\begin{equation}
f(x)=\frac{1}{1-x}, \qquad (x\in\mathbb{C}, \quad x\neq 1).
\label{exact}
\end{equation} 

In mathematical manipulations of physical equations, 
there is a physical reason for analytic continuation. 
Due to the use of certain computational techniques 
or one's choice of formulation, 
the validity of some mathematical expressions may be restricted, 
but often the physical quantities we are computing 
could be well-defined with a larger range of validity.
Relying on the analyticity of the physical problem, 
analytic continuation allows us to retrieve the full range of validity
of our results, even though the validity of derivation is more restricted. 

As an example, imagine that in a physical problem, 
we need to solve the following differential equation
\begin{equation}
(1-x)f'(x)-f(x)=0.
\end{equation}
One might try to solve this differential equation as an expansion 
\begin{equation}
f(x)=f_0+f_1x+f_2x^2+\cdots,
\end{equation}
and obtain some recursion relations which results in the solution (\ref{seriessol}), 
up to an overall constant. 
If one analytically continues this result to (\ref{exact}), 
one can directly check that it is the correct solution of the differential equation 
even for $x$ outside the range $(-1,1)$.
The appearance of the series (\ref{seriessol}) 
and the convergence condition $|x| < 1$ is merely an artifact of 
the technique used in derivation.

\subsection{One-loop diagrams in string theory}

In this subsection, 
we shall review how UV divergence is avoided in string theory 
via analytic continuation.

Apart from factors involving vertex operators, 
the formula for the amplitudes of one-loop diagrams 
contain a common factor \cite{6}
\begin{equation}
A_0=\int_0^1d\omega \; \Tr \omega^{L_0-2},
\label{openstring1}
\end{equation}
where $L_0=\frac12p^2+N$.
This factor comes from the self energy diagram of an open string.
\begin{figure}[h]
\begin{center}
\includegraphics[]{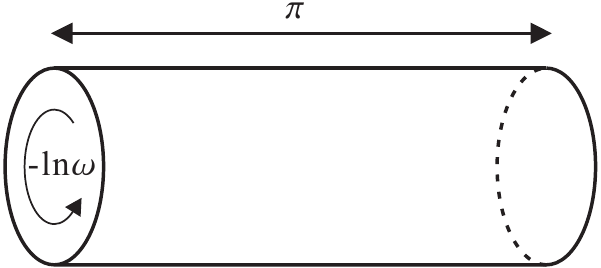}
\end{center}
\caption{}
\label{worldsheet1}
\end{figure}
The trace of eq. (\ref{openstring1}) 
includes summation over each state in the spectrum 
and integration of energy-momentum. 
The factor $\omega^{L_0}$ is an operator that 
propagates a string through a proper time of length $(-\ln\omega)$ 
(which is positive, since $\omega\le1$). 
The regime $\omega\rightarrow 1$ corresponds to a very short proper time, 
and thus a very narrow cylinder (see Fig.\ref{worldsheet1}); 
this is thus the ultraviolet regime. 

To take a closer look at the UV behavior of $A_0$ (\ref{openstring1}), 
one can formally compute $A_0$ as 
\footnote{
For bosonic strings, the term $0^{L_0-1}$ also diverges. 
But our attention is on the UV divergent terms 
due to integration over the regime $\omega \sim 1$.
}
\bea
A_0&=&\Tr\frac{1}{L_0-1}\omega^{L_0-1}|_0^1 \nn \\
&=&\int d^D p\;\sum_N \frac{c_N}{p^2+N-1}(1^{L_0-1}-0^{L_0-1}) \nn \\
&=&\int d^D p\;\sum_N \frac{c_N}{p^2+N-1}.
\label{stringdivergence}
\eea
The factor $c_N$ comes from the symmetry factor of particles 
depending on their spin.
Here we notice that 
the momentum integration leads to a UV divergence 
for each particle propagator. 
Naively, the sum over the contributions from infinitely many particles 
can only make the UV divergence infinitely worse.
But it is well known that string theory is free from UV divergence. 
We will see below that the trick is analogous to the analytic continuation 
we used to regularize the scalar field theories.

\begin{figure}
\begin{center}
\includegraphics[scale=0.67]{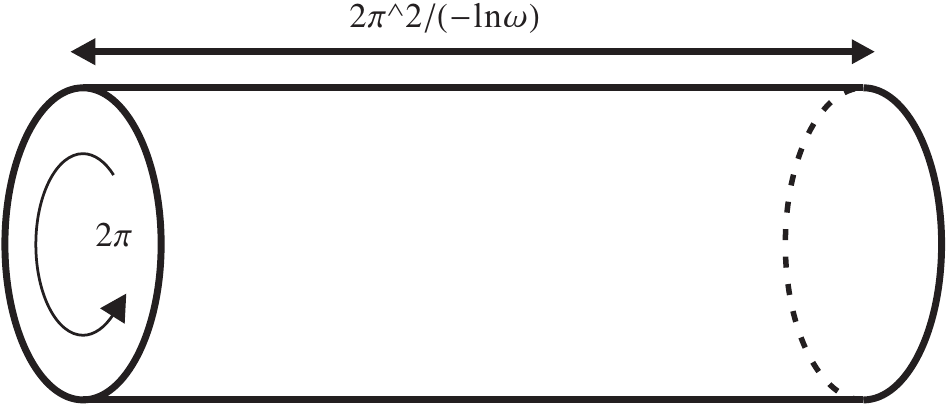}
\end{center}
\caption{}
\label{worldsheet2}
\end{figure}

Let us recall that string theory solves this UV problem 
by conformal symmetry and open-closed string duality. 
By scaling symmetry, 
a cylinder with length $\frac{2\pi^2}{-\ln\omega}$ (see Fig. \ref{worldsheet2})
and circumference $2\pi$ is equivalent to Fig. \ref{worldsheet1}. 
We can look at this diagram with a different perspective, 
interpreting it as a propagating closed string over a proper time 
$\frac{2\pi^2}{-\ln\omega}$. 
When $\omega$ is close to 1, the cylinder is very long. 
The $\omega\rightarrow 1$ regime is no longer the ultraviolet regime,
but the infrared regime. 

Defining $-\ln q\equiv\frac{2\pi^2}{-\ln\omega}$, 
let us consider a closed string propagating for the proper time $(-\ln q)$. 
The amplitude for this process is of the form
\begin{equation}
A_0\sim\int_0\frac{dq}{q}\; U,
\label{closedoneloop}
\end{equation}
where $U$ is the evolution operator that carries the initial state to the final state
\begin{equation}
U=\langle f | e^{(L_0+\tilde{L_0}-2)\ln q} | i \rangle.
\end{equation}
In the infrared regime $\omega\rightarrow 1$,
a closed string with zero momentum propagates for a very long proper time. 
It follows that the states $|i\rangle$ and $|f\rangle$ are of nearly zero momentum, 
and classical trajectories dominate the path integral. 
Since the equation of motion and level matching condition 
$L_0=\tilde{L}_0$ are satisfied for the classical trajectories, 
we expect that
\begin{equation}
U\sim\langle f | 0\rangle
\langle 0e^{(2L_0-2)\ln q} |0\rangle
\langle 0|i\rangle.
\end{equation} 
As there is a $1/q$ factor in the denominator in (\ref{closedoneloop}), 
the only divergence in $A_0$ comes from the tachyon ($L_0=0$), 
which can be removed in superstring theories, 
and the infrared divergence of the dilaton ($L_0=1$), 
which is analogous to the IR divergence typical in massless field theories.
Therefore, one concludes that the UV divergence
in an open superstring one-loop diagram disappears 
if we compute it in the closed string picture.
Since it is the string worldsheet duality that allows us to identify 
the ill-defined expansion of $A_0$ in (\ref{stringdivergence})
with the UV-finite closed string tree level diagram,
we want to take a closer look at the string worldsheet duality 
and compare it with our trick of analytic continuation.
%Of course there is no open-closed string duality in an ordinary field theory, 
%but this gives us a hint. 
%When a duality exists, 
%the same physical quantity can be expanded in two different ways. 
%If one expansion is ill-defined, perhaps this only means that 
%we used a bad expansion, and we can just try the other expansion. 
%The basic idea is therefore the same as analytic continuation. 

\subsection{$s-t$ duality and analytic continuation}

The simplest example of string worldsheet duality is 
the $s-t$ duality of open-string 4-point amplitudes.
Consider the 4-point tree-level scattering amplitudes in $t-$channel 
\begin{equation}
A(s,t)=-\sum_J\frac{g^2_J(-s)^J}{t-M^2_J},
\label{eqn41}
\end{equation}
and $s-$channel
\begin{equation}
A'(s,t)=-\sum_J\frac{g^2_J(-t)^J}{s-M^2_J}.
\label{eqn42}
\end{equation} 
Note that in order for the two quantities to be identical $A(s, t) = A'(s, t)$, 
the sum $\sum_J$ must be an infinite series
and the masses $M_J$ and couplings $g_J$ must be fine-tuned.
The parameters are chosen such that
\begin{subequations}
\begin{align}
A(s,t)&=-\sum_{n=0}^{\infty}
\frac{(\a(s)+1)(\a(s)+2)\cdots(\a(s)+n)}{n!}\frac{1}{(\a(t)-n)},\\
A'(s,t)&=-\sum_{n=0}^{\infty}
\frac{(\a(t)+1)(\a(t)+2)\cdots(\a(t)+n)}{n!}\frac{1}{(\a(s)-n)},
\end{align}
\end{subequations}
where $\a(s) = \a' s + \a_0$.
Since $\a'>0$, when $s>0$ and $t>0$, by the relation of gamma function
\begin{equation}
\Gamma(n+1)=n\Gamma(n),
\end{equation}
the amplitudes can also be written as
\begin{subequations}
\begin{align}
A(s,t)&=-\sum_{n=0}^{\infty}
\frac{\Gamma(\a(s)+n+1)}{\Gamma(\a(s)+1)n!}\frac{1}{(\a(t)-n)},\\
A'(s,t)&=-\sum_{n=0}^{\infty}
\frac{\Gamma(\a(t)+n+1)}{\Gamma(\a(t)+1)n!}\frac{1}{(\a(s)-n)}.
\end{align}
\label{eqn45}
\end{subequations}

These expressions seem divergent at first sight. 
According to Stirling's formula, 
the numerator of each term is of order $n^{\a(s)}$ or $n^{\a(t)}$ 
and the denominator is of order $1/n$.
But if we first assume that $\a' < 0$, 
the series (\ref{eqn45}) converge to the form of an expansion of the beta function
\begin{equation}
B(x,y)=\sum_{n=0}^\infty\frac{\Gamma(n-y+1)}{n!\Gamma(-y+1)}
\frac{1}{x+n},\qquad y>0.
\end{equation}
Then we can analytically continue the quantities back to $\a'>0$, 
and see that  
both $A(s,t)$ and $A'(s,t)$ in (\ref{eqn45}) can be expressed 
by the well-known Veneziano amplitude
\begin{equation}
A(s,t)=\frac{\Gamma(-\a(s))\Gamma(-\a(t))}{\Gamma(-\a(s)-\a(t))}.
\end{equation}

In the sample calculation above, 
we reminded ourselves that the worldsheet duality, 
which is at the heart of UV-finiteness of string theory, 
is also a result of analytic continuation --
the same trick we used to remove UV divergences in our field theory models. 
The duality that interchanges 
a one-loop open string diagram with a tree-level closed string diagram 
is a result of the Wick rotation on the worldsheet. 
The Wick rotation is an analytic continuation. 
The infinite number of poles, fine-tuned masses and couplings 
in string theory are all reminiscent to 
our choice of the propagator (\ref{propagator}).
The key ingredients that allow us to remove UV divergences 
are exactly the same in our model and in string theory. 
The only difference is that in string theory the (much finer) fine-tuning 
leads to a large symmetry (conformal symmetry),
and is capable of removing UV divergences 
even in the presence of vector and tensor fields. 
It is tempting to make the conjecture that 
the fine-tuning conditions (\ref{generalcond}) 
also correspond to some symmetries. 
We leave this question for future investigation.

\section{Conclusion}

Let us summarize our results. 
For a scalar field theory in $d$-dimensional spacetime
($d$ must be even)
with an action of the form 
\be
\label{S1}
S = \int d^d x \; \left( \frac{1}{2} \phi f^{-1}(-\del^2) \phi 
- V(\phi) \right), 
\ee
where $V(\phi)$ is a polynomial of $\phi$ of arbitrary order, 
and the function $f(-\del^2)$ in the kinetic term is given in (\ref{propagator}), 
with the conditions in (\ref{generalcond}) satisfied, 
the theory is {\em UV-finite, unitary and Lorentz invariant}
to all orders in the perturbative expansion. 
Remarkably, the conditions (\ref{generalcond}) 
are independent of the order of the polynomial interactions.
It should be straightforward to generalize our discussion above 
to scalar field theories (\ref{S1}) with more than one scalar fields $\phi_a$ 
with polynomial type interactions.

The prescription for calculating Feynman diagrams is 
to first use dimensional regularization to regularize 
integrals over internal momenta, 
and then impose the conditions (\ref{generalcond}) 
to remove all UV divergences in the limit $\epsilon \rightarrow 0$.
The infinite sums involved in the calculation are dealt with via 
analytic continuation. 
Roughly speaking, 
the conditions (\ref{generalcond}) remove  
the first $d/2$ terms in the large $k$ expansion of the propagator
\be
f(k^2) = \sum_n \frac{c_n}{k^2+m_n^2} = 
\frac{\sum_n c_n}{k^2} - \frac{\sum_n c_n m_n^2}{k^4}
+ \frac{\sum_n c_n m_n^4}{k^6} - \cdots.
\ee
Hence the propagator goes to zero as fast as $1/k^{d+2}$
as $k \rightarrow \infty$, 
removing UV divergences for all diagrams.

Since the propagator $f(k^2)$ is the same as the sum over 
ordinary propagators of particles of mass $m_n$ 
with a normalization constat $c_n$, 
the perturbation theory of (\ref{S1}) is the same as that of the action
\be
\label{S2}
S' = \int d^d x \; \left( \sum_n \frac{1}{2c_n} \phi_n (-\del^2 + m_n^2) \phi_n
- V({\sum_n \phi_n}) \right).
\ee
The same action can also be written as 
\be
\label{S3}
S'' = \int d^d x \; \left( 
\sum_n \frac{1}{2} \hat{\phi}_n (-\del^2 + m_n^2) \hat{\phi}_n
- V({\sum_n \sqrt{c_n} \hat{\phi}_n}) \right), 
\ee
where $\hat{\phi}_n = \phi_n/\sqrt{c_n}$.
Therefore, the nonlocal scalar field theory (\ref{S1}) is equivalent to 
a theory of infinitely many scalar fields with fine tuned masses 
and coupling constants.

The analogy between string theory and the higher derivative theory 
defined by the propagator (\ref{propagator}) 
was made in Sec. \ref{AC}. 
While the worldsheet conformal symmetry justifies 
the analytic continuation and fine tuning of the mass spectrum 
in string theory, 
it would be of crucial importance to search for 
a symmetry principle underlying 
the fine-tuning conditions (\ref{generalcond}).
We notice that the partition function 
\be
Z[J_n] = \int \prod_n D\hat{\phi}_n \; 
e^{-S''[\hat{\phi}_n]+\int d^d x \; \hat{\phi}_n J_n}
\ee
has the algebraic property
\be
Z[J_n + \alpha \sqrt{c_n}m_n^{2r+2}] = 
e^{\int dx^d \; \sum_n (\frac{1}{2} \alpha^2 c_n m_n^{4r+2}
+ \alpha \sqrt{c_n} m_n^{2r} J_n)} 
Z[J_n] 
\qquad 
(r = 0, 1, \cdots, (d-2)/2). 
\ee
which implies that the quantity
\be
\tilde{Z}[J_n] \equiv 
e^{-\int d^d x \; \sum_n \frac{1}{2m_n^2} J_n^2} Z[J_n]
\ee
is invariant under the transformation 
\be
J_n \rightarrow J_n + \alpha \sqrt{c_n} m_n^{2r+2} \qquad
(r = 0, 1, \cdots, (d-2)/2).
\ee
However, 
the physical significance of this algebraic property 
and the underlying symmetry principle 
still remain mysterious. 

Another direction for future study is to extend our results to 
field theories of various spins. 
It will be very interesting to generalize 
our approach to incorporate gauge fields, the graviton, 
and even higher spin fields.

\subsection*{Acknowledgments}

The authors thank Chuang-Tsung Chan, Chien-Ho Chen, Ru-Chuen Hou, 
Yu-Ting Huang, Takeo Inami, Hsien-Chung Kao, Yeong-Chuan Kao, 
Yutaka Matsuo, Darren Sheng-Yu Shih and Chi-Hsien Yeh 
for helpful discussions. 
This work is supported in part by the National Science Council,
and the National Center for Theoretical Sciences, Taiwan, R.O.C.

\end{document}